\documentclass[useAMS,usenatbib]{mn2e}
\usepackage{epsfig}
\usepackage{amsmath, amssymb,bm}
\usepackage[varg]{txfonts}

\usepackage{color}

\title[The GM disc instability with a viscous dead zone]{The
  gravo-magneto disc instability with a viscous dead zone}

\author[R. G. Martin \& S. H. Lubow]{Rebecca
  G. Martin$^{1,3}$\thanks{E-mail: rebecca.martin@jila.colorado.edu}
  and Stephen H. Lubow$^2$\\ $^1$JILA, University of Colorado,
  Boulder, CO 80309, USA\\ $^2$Space Telescope Science Institute, 3700
  San Martin Drive, Baltimore, MD 21218, USA \\ $^3$NASA Sagan Fellow \\}

\begin{document}

\date{}

\pagerange{\pageref{firstpage}--\pageref{lastpage}} 
\pubyear{2013}
\maketitle

\label{firstpage}

\begin{abstract}
We consider the evolution of accretion discs that contain some
turbulence within a disc dead zone, a region about the disc midplane
of a disc that is not sufficiently ionised for the magneto-rotational
instability (MRI) to drive turbulence.  In particular, we determine
whether additional sources of turbulence within a dead zone are
capable of suppressing gravo-magneto (GM) disc outbursts that arise
from a rapid transition from gravitationally produced to MRI produced
turbulence.  With viscous $\alpha$ disc models we consider two
mechanisms that may drive turbulence within the dead zone. First, we
examine a constant $\alpha$ parameter within the dead zone. This may
be applicable to hydrodynamical instability, such as baroclinic
instability, where the turbulence level is independent of the MRI
active surface layer properties. In this case, we find that the disc
will not become stable to the outbursts unless the dead zone turbulent
viscosity is comparable to that in the MRI active surface layers.
Under such conditions, the accretion rate through the dead zone must
be larger than that through the MRI active layers. In a second model,
we assume that the accretion flow though the dead zone is a constant
fraction (less than unity) of that through the active layers.  This
may be applicable to turbulence driven by hydrodynamic waves that are
excited by the MRI active surface layers. We find that the instability
is hardly affected by the viscous dead zone. In both cases however, we
find that the triggering of the MRI during the outburst may be due to
the heating from the viscosity in the dead zone, rather than
self-gravity. Thus, neither mechanism for generating turbulence within
the dead zone can significantly stabilise a disc or the resulting
outburst behaviour.
\end{abstract}


\section{Introduction}

Accretion discs transport angular momentum outwards allowing material
to accrete on to the central object \citep{pringle81}. In a fully
ionised disc, the gas is well coupled to the magnetic field and the
magneto rotational instability (MRI) drives turbulence and thus
angular momentum transport \citep{balbus91}. However, protoplanetary
discs are thought to have a sufficiently low ionisation fraction that
a dead zone forms about the midplane where the MRI cannot efficiently
transport angular momentum \citep{gammie96,gammie98, turner08}. When a
dead zone forms, for a range of accretion rates, the disc becomes
unstable to the gravo-magneto (GM) disc instability, where the
turbulence cycles between magnetic and gravitational
\citep{armitage01, zhu09, zhu10a, zhu10b, martin11a, martin13}.

Consider the case that a dead zone has zero viscosity.  In that
situation, a steady state disc cannot contain a dead zone.  Instead,
there are three possible steady state solutions at each radius in the
disc. First, if the midplane temperature is sufficiently high there is
a thermally ionised MRI steady state. Secondly, if the surface density
is sufficiently low there is an externally ionised MRI steady
state. The surface layers of the disc may be ionised by external
sources such as cosmic rays or X-rays from the central star
\citep{glassgold04}.  Turbulence in the dead zone may be driven by
self-gravity if the disc becomes sufficiently massive
\citep{paczynski78,lodato04} and this can drive the third steady state
which has a self gravitating dead zone with MRI active surface layers.
However, there may be a range of radii in the disc for which there
exists no steady state and then the disc is unstable to the GM disc
instability \citep{martin11a, martin13}.

We have explained the GM disc instability as transitions between these
steady state solutions by plotting a state diagram of the accretion
rate through the disc against the surface density for a fixed radius
\citep{martin11a,martin13}. This is similar to the thermal-viscous
instability ``S-curve'' used to explain dwarf nova outbursts
\citep{bath82,faulkner83}. In our previous work on the gravo-magneto
disc instability we assumed that the viscosity in the dead zone is
zero \citep{martin11a,martin12a,martin12b,martin13} but this is
somewhat uncertain.

There are at least two mechanisms that may drive turbulence in the
dead zone. First, it is possible that turbulence within the dead zone
may be driven by hydrodynamic instabilities such as the baroclinic
instability \cite[e.g.][]{klahr03,petersen07,lesur10}.  In a
baroclinic state, the pressure varies over surfaces of constant
density.  The non-axisymmetric misalignment between surfaces of
constant density and surfaces of constant pressure generates
vorticity. Vortices within a dead zone could be associated with a
viscosity parameter as high as $\alpha=5\times 10^{-3}$
\citep{lyra11}. However, there is still some uncertainty over the
viability of such a mechanism for driving turbulence within a
protoplanetary disc. The conditions under which the baroclinic
instbility can drive turbulence are not well constrained \citep[see
  e.g.][]{armitage11}. The dissipation from the baroclinic instability
is poorly known as vortices behave very differently to turbulence. To
model the effects of such hydrodynamic instabilities, we take a
constant $\alpha$-parameter in the dead zone of a layered disc model.

Secondly, shearing box simulations suggest that the MHD turbulence
generated in the disc surface layers may produce some hydrodynamic
turbulence in the dead zone layer that may produce a small but
non-zero viscosity \citep[e.g.][]{fleming03,simon11,gressel12}.
Hydrodynamic waves are excited by the turbulence in the active surface
layers. These can penetrate to the midplane and exert a nonzero
Reynolds stress there. The vertically stratified local simulations of
\cite{fleming03} find the midplane Reynolds stress in the dead zone to
be less than an order of magnitude below the Maxwell stress in a fully
turbulent disc. However, those simulations used an isothermal equation
of state, and the strength of Reynolds stresses communicated to the
midplane may be sensitive to that assumption and to the thickness of
the active surface layer \citep{bai09,armitage11}. Thus, there is
still some uncertainty associated with the strength of such
turbulence. In this work, we model the turbulence driven in the dead
zone from the active layer with an $\alpha$-parameter that varies such
that the accretion rate through the dead zone is a constant fraction
of that through the active layers.

\cite{bae13} considered the time-dependent evolution of a disc with a
small viscosity in the dead zone that is gravo-magneto unstable. In
this work we consider how large the viscosity in the dead zone can be
before the disc reaches a fourth steady state, a viscous dead zone
steady state, thus stabilising the outburst behaviour. We consider a
protoplanetary disc model but we note that the gravo-magneto disc
instability can occur in discs on a range of scales. For example, dead
zones and the GM instability also likely occur in circumplanetary
discs \citep{martin11b,lubow12,lubow13}.

In Section~\ref{model} we describe the layered disc model. In
Section~\ref{sec:ad} we present the results for disc solutions with a
constant viscosity in the dead zone. These may be appropriate for a
dead zone viscosity driven by the baroclinic instability, for
example. In Section~\ref{sec:md} we consider solutions with an
accretion rate through the dead zone that is a fixed fraction of that
through the active layers. These may be more appropriate for dead zone
viscosities driven from the turbulence in the active layers.

\section{Layered Disc Model}
\label{model}

\begin{figure*}
\includegraphics[width=7cm]{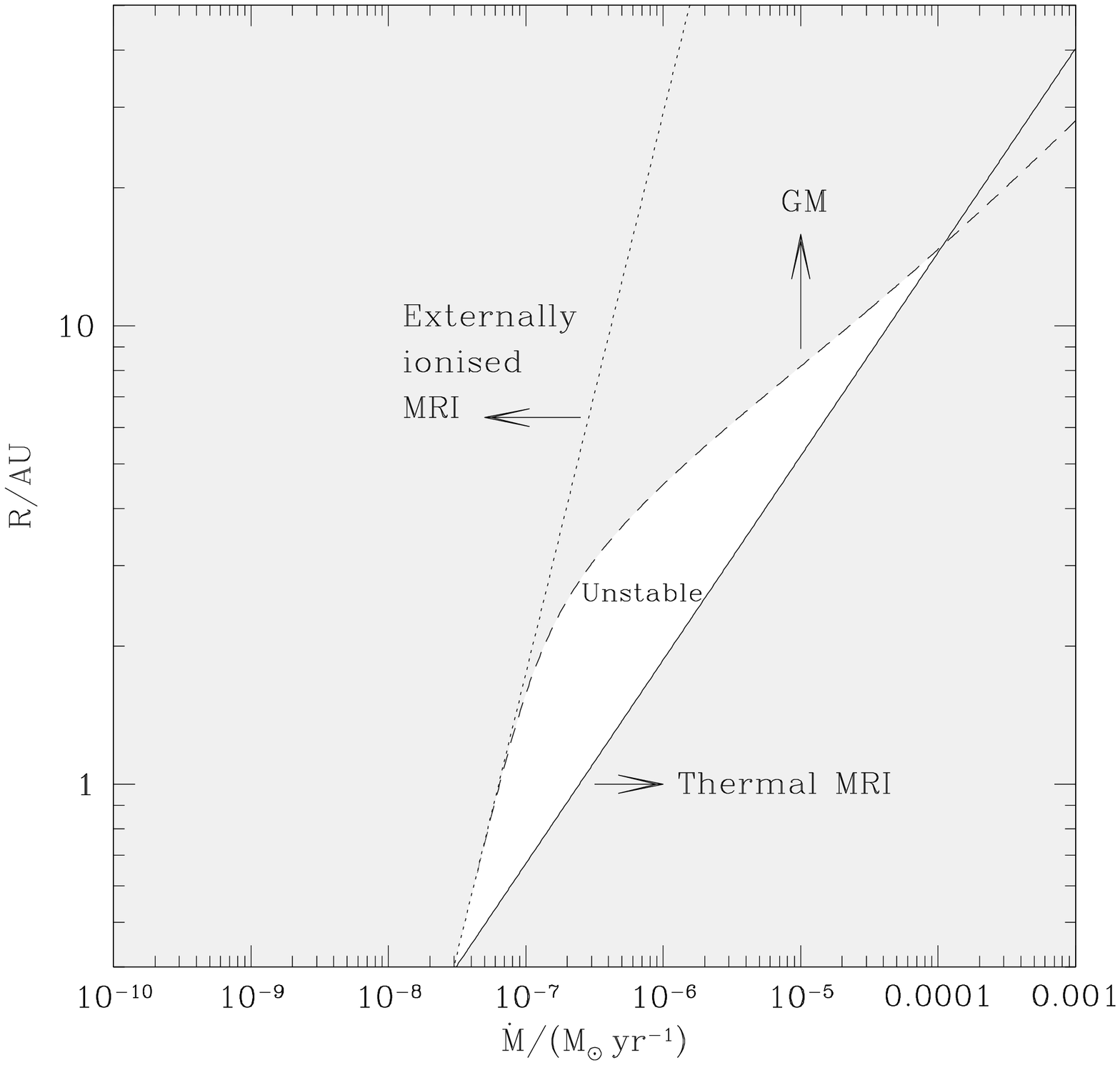}
\includegraphics[width=7cm]{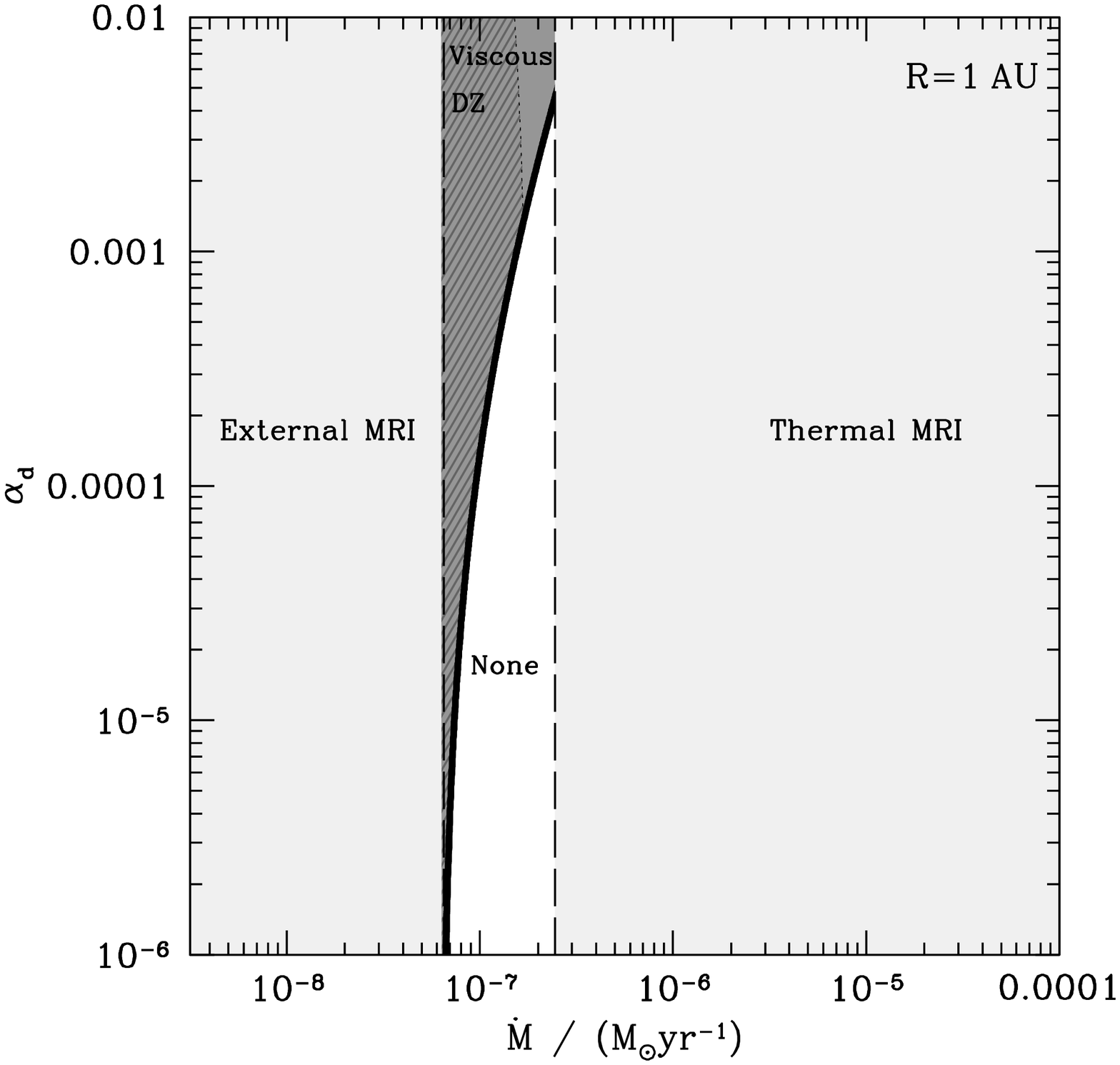}
\includegraphics[width=7cm]{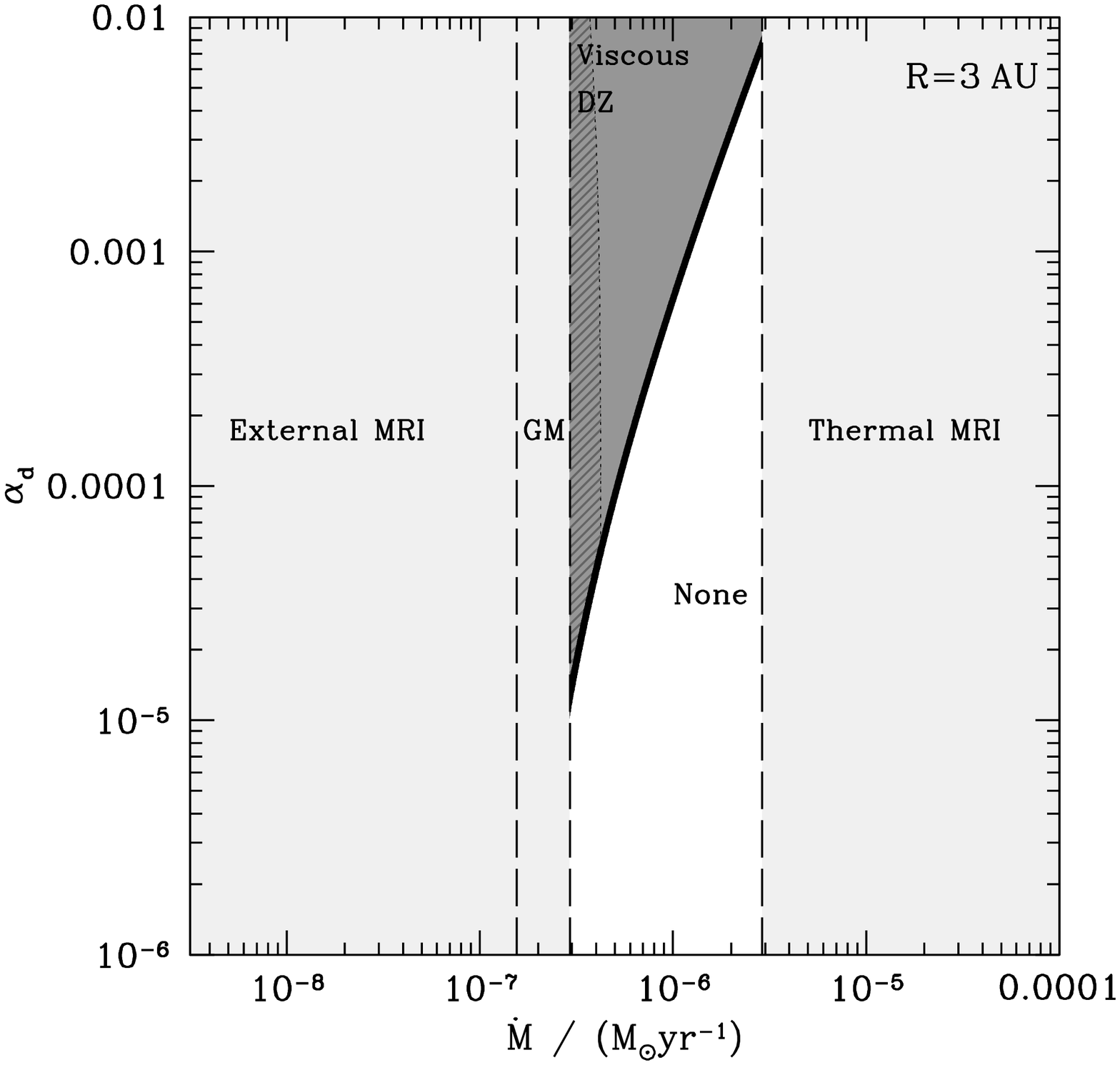}
\includegraphics[width=7cm]{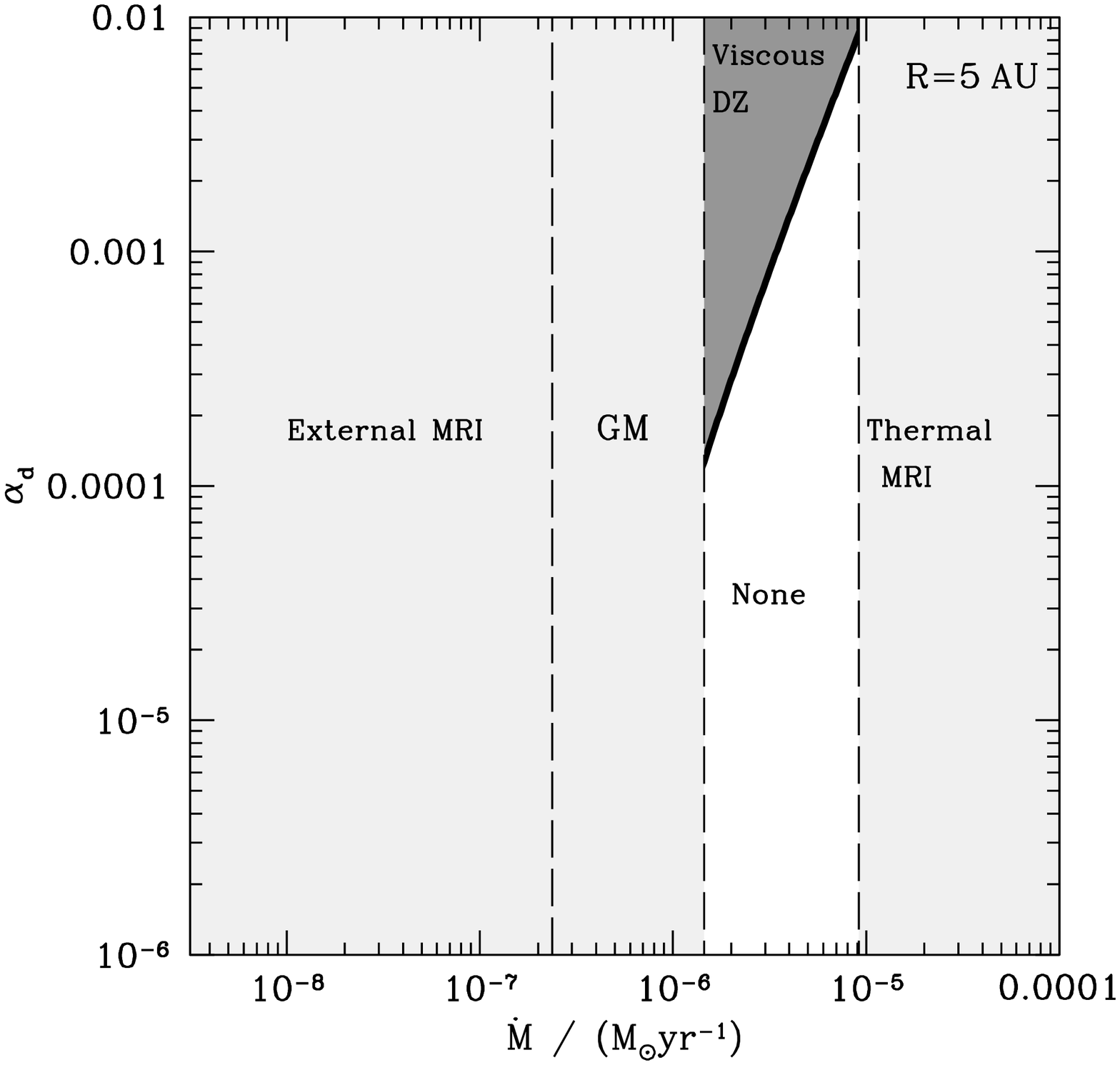}
\caption{A disc with $\Sigma_{\rm crit}=200\,\rm g\,cm^{-2}$, $T_{\rm
    crit}=800\,\rm K$ and $\alpha_{\rm m}=0.01$. Top Left: Radii in
  the disc for which steady state MRI and GM disc solutions exist for
  $\alpha_{\rm d}=0$ for varying infall accretion rate. The shaded
  region shows where such a steady state exists. The unshaded region
  has no steady state and thus if the accretion rate lies within this
  region the disc may be GM unstable.  The remaining three plots show
  the steady solutions including the viscous dead zone solution for
  varying dead zone viscosity for $R=1\,\rm AU$ (top right), $R=3\,\rm
  AU$ (bottom left) and $R=5\,\rm AU$ (bottom right). The steady
  thermally and externally ionised MRI solutions and GM solutions
  exist for accretion rates in the pale shaded regions and the long
  dashed vertical lines denote the boundaries between these
  regions. In the dark shaded region there is a viscous dead zone
  steady state with viscosity parameter $\alpha_{\rm d}$. In the dark
  hatched region (where it exists), the mass flow through the dead
  zone layer is less than that through the active layer ($\dot M_{\rm
    d}<\dot M_{\rm m}$). Similarly, in the dark unhatched region, the
  accretion rate through the dead zone is greater than that through
  the active layer ($\dot M_{\rm d}>\dot M_{\rm m}$). }
\label{data}
\end{figure*}

\begin{figure*}
\includegraphics[width=7cm]{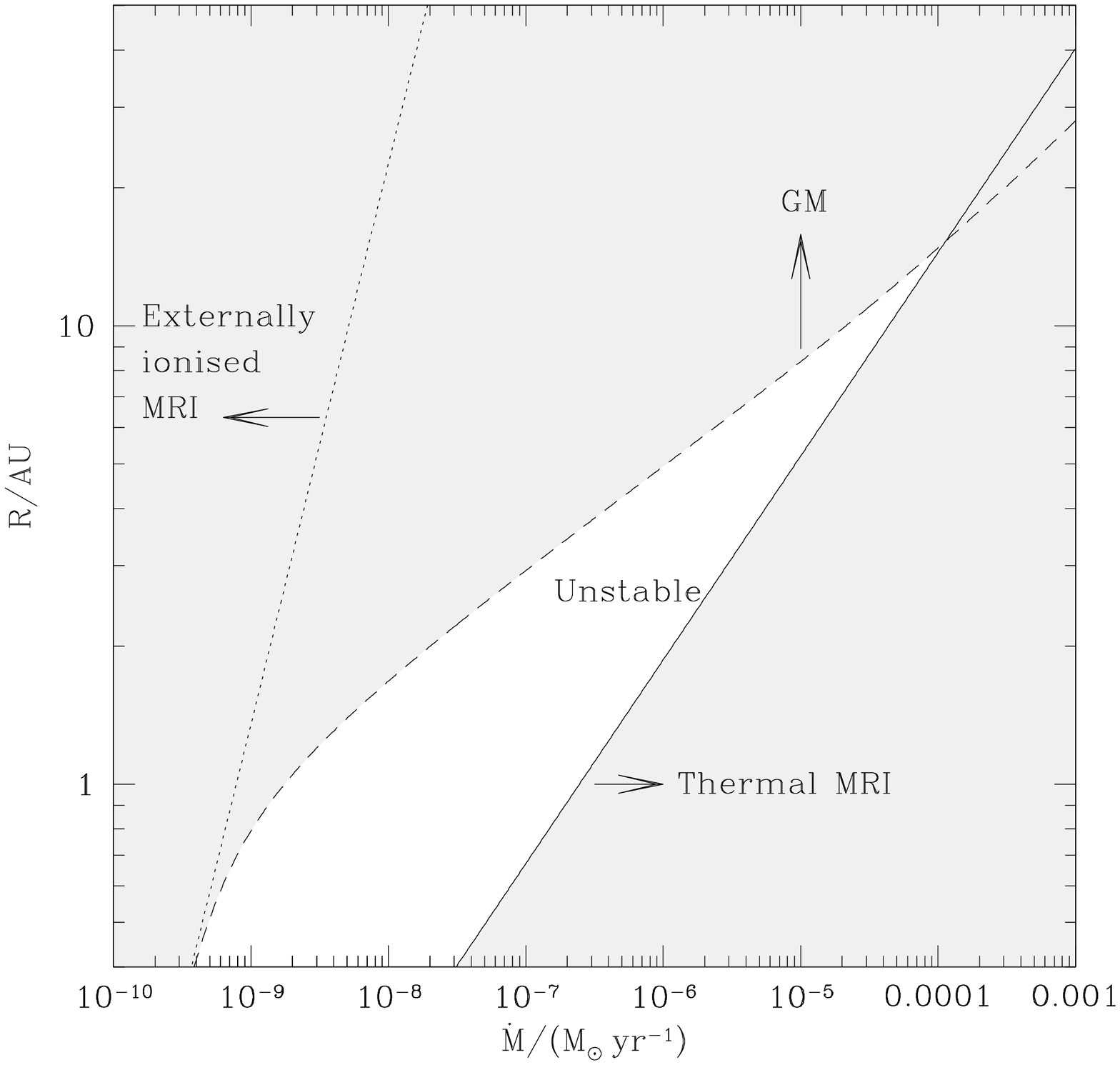}
\includegraphics[width=7cm]{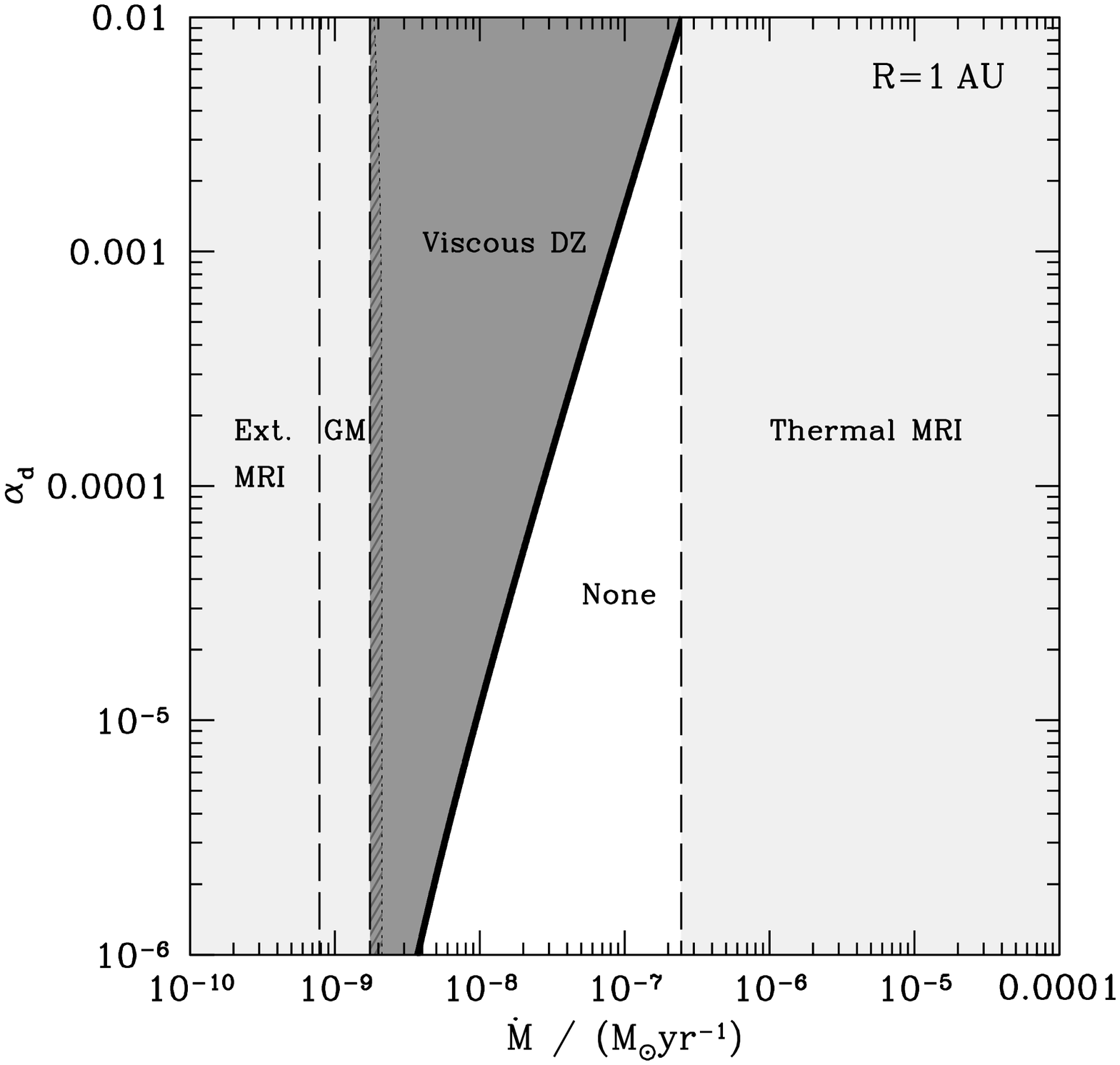}
\includegraphics[width=7cm]{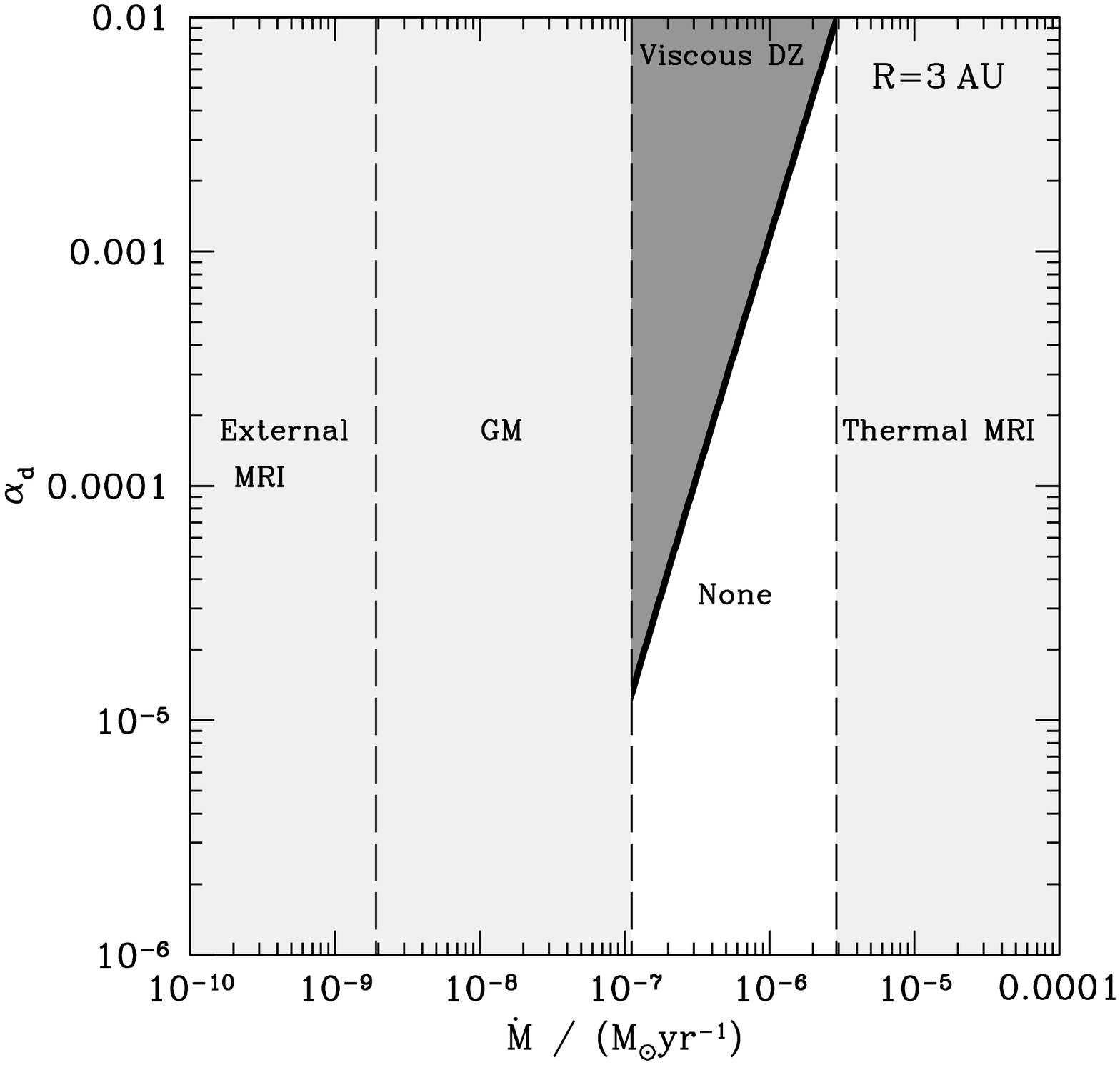}
\includegraphics[width=7cm]{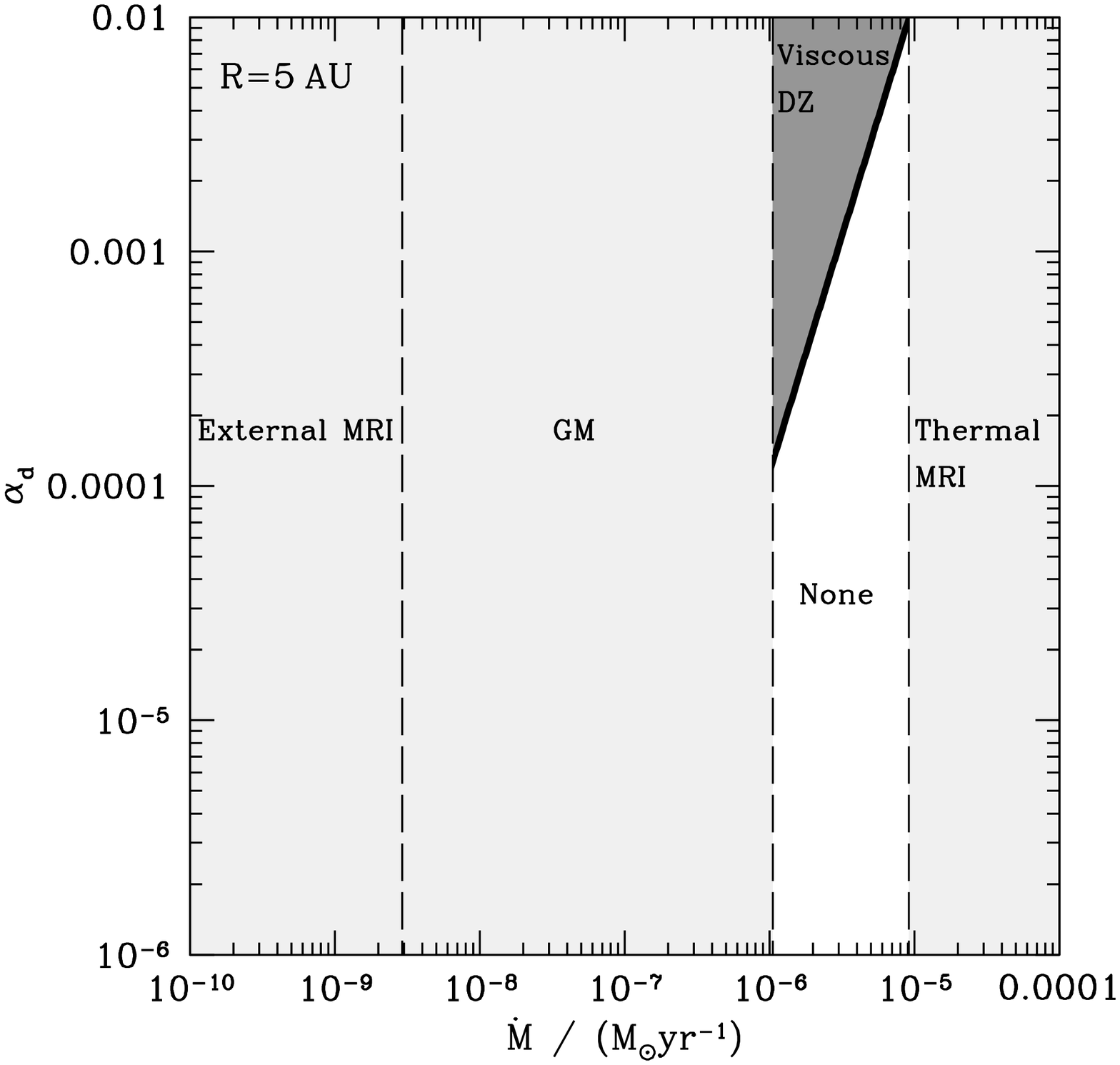}
\caption{Same as Fig.~\ref{data} but with $\Sigma_{\rm crit}=20\,\rm
  g\,cm^{-2}$. The dark shaded hatched region for the viscous dead
  zone solutions with $\dot M_{\rm d}<\dot M_{\rm m}$ exists only in a
  very small region of parameter space at $R=1\,\rm AU$ and does not
  exist for $R=3$ or $R=5\,\rm AU$..  }
\label{data2}
\end{figure*}

We follow the disc model of \cite{martin11a} that was initially
developed by \cite{armitage01}.  Material in the disc orbits the
central star of mass $M=1\,\rm M_\odot$ at radius $R$ with Keplerian
velocity, $\Omega=\sqrt{G M/R^3}$.  If the temperature of the disc is
greater than the critical value, $T>T_{\rm crit}$, then the disc is
thermally ionised and fully MRI active. The value of the critical
temperature is thought to be around $T_{\rm crit}=800\,\rm K$
\citep{umebayashi83}.  Similarly if the surface density of the disc is
smaller than the critical that can be ionised by external sources
(such as cosmic rays and X-rays), $\Sigma<\Sigma_{\rm crit}$, the disc
has a fully MRI active solution. However, if neither of these
conditions are satisfied, then there is a dead zone layer about the
midplane.

There remains some uncertainty in the critical surface density that
may be ionised by external sources, $\Sigma_{\rm crit}$. If cosmic
rays are the dominant ionisation source, $\Sigma_{\rm crit}\approx
200\,\rm g\,cm^{-2}$ \citep{gammie96,fromang02}. However, without
cosmic rays, X-rays from the central star may dominate and in this
case the active layer surface density is much smaller
\citep{matsumura03}.  Ionisation balance calculations determine
$\Sigma_{\rm crit}$ in a disc subject to various processes such as
ambipolar diffusion and the presence of polycyclic aromatic
hydrocarbon and dust and find that they suppress the ionisation
fraction further \citep[e.g.][]{bai09,perez11,bai11,simon13}.
However, these calculations predict accretion rates much lower than
those observed in typical T Tauri stars that suggest $\Sigma_{\rm
  crit}>10\,\rm g\,cm^{-2}$ \citep{perez11}. In this work, we consider
two values, $\Sigma_{\rm crit}=200\,\rm g\,cm^{-2}$ and $\Sigma_{\rm
  crit}=20\,\rm g\,cm ^{-2}$.

To find the steady state disc solutions we solve the accretion disc
equation
\begin{equation}
\dot M=3 \pi \left( \nu_{\rm m} \Sigma_{\rm m}+\nu_{\rm d}\Sigma_{\rm d}\right)
\label{mdot}
\end{equation}
\citep{pringle81}, where $\dot M$ is the steady accretion rate through
the disc, $\Sigma_{\rm m}$ is the surface density in the MRI active
surface layers, $\Sigma_{\rm d}$ is the surface density in the dead
zone layer, the total surface density is $\Sigma=\Sigma_{\rm
  m}+\Sigma_{\rm d}$. The accretion rate through the MRI active
surface layers is $\dot M_{\rm m}=3\pi \nu_{\rm m}\Sigma_{\rm m}$ and
the accretion rate through the dead zone is $\dot M_{\rm
  d}=3\pi \nu_{\rm d} \Sigma_{\rm d}$. The viscosity in the MRI active
surface layers is
\begin{equation}
\nu_{\rm m}=\frac{\alpha_{\rm m}c_{\rm m}^2}{\Omega},
\end{equation}
where $c_{\rm m}=\sqrt{kT_{\rm m}/\mu}$ is the sound speed, $T_{\rm m}$
is the temperature and $\alpha_{\rm m}$ is the \cite{shakura73}
viscosity parameter that we take to be $\alpha_{\rm m}=0.01$
\citep[see e.g.][]{hartmann98}.

The dead zone may become self gravitating if the \cite{toomre64}
parameter becomes less than the critical,
\begin{equation}
Q=\frac{c_{\rm d}\Omega}{\pi G \Sigma}<Q_{\rm crit},
\end{equation}
where $c_{\rm d}=\sqrt{kT_{\rm c}/\mu}$ is the sound speed in the midplane
layer of temperature $T_{\rm c}$ and we take $Q_{\rm crit}=2$.  The dead zone
layer has viscosity
\begin{equation}
\nu_{\rm d} =\left(\alpha_{\rm d}+\alpha_{\rm g} \right)\frac{c_{\rm d}^2}{\Omega},
\end{equation}
where the part due to self gravity is
\begin{equation}
\alpha_{\rm g}=\alpha_{\rm m}\left[\left(\frac{Q_{\rm crit}}{Q}\right)^2-1\right]
\label{ag}
\end{equation}
if $Q<Q_{\rm crit}$ and zero otherwise \citep{lin87,lin90}. The form
of the $Q$ dependence in equation~(\ref{ag}) does not affect the disc
evolution provided it is a strongly decreasing function \citep[see
  also][]{zhu10b,martin13}. In fact, we find that the addition of the
residual dead zone viscosity term, $\alpha_{\rm d}$, also does not
affect the viscosity in the dead zone, if the disc is self gravitating
\citep[unless it is close to the fully MRI active value, $\alpha_{\rm
    d}\approx \alpha_{\rm m}$, see the discussion section in][for more
  details]{martin13}. We consider two different models for
$\alpha_{\rm d}$. First we consider $\alpha_{\rm d}=\,\rm constant$
(see Section~\ref{sec:ad}) and then we consider $\alpha_{\rm d}$ that
satisfies $\dot M_{\rm d}=f \dot M_{\rm m}$, where $f=\,\rm constant$
(see Section~\ref{sec:md}). All solutions with a dead zone have MRI
active surface layers with surface density $\Sigma_{\rm m}=\Sigma_{\rm
  crit}$ that are ionised by the external sources.

We also consider the restriction that the accretion rate through the
dead zone, $\dot M_{\rm d}$, is less than the accretion rate through
the active layer, $\dot M_{\rm m}$.  If the dead zone viscosity is
driven from the active layer, it is unlikely that a small amount of
turbulence in the surface layers can drive a large amount of
turbulence in a more massive dead zone region \citep{zhu09}. On the
other hand, if the baroclinic instability is driving the turbulence,
it is possible that the viscosity in the dead zone may be higher than
that in the active layers because it is not determined by the
turbulence in the surface layers.

The mass conservation equation is solved coupled with a steady energy
equation for the surface temperature, $T_{\rm e}$,
\begin{equation}
\sigma T_{\rm e}^4 = \frac{9}{8} \Omega^2 \left( \nu_{\rm
  m}\Sigma_{\rm m} + \nu_{\rm d}\Sigma_{\rm d} \right)
\label{energy}
\end{equation}
\citep{pringle86, cannizzo93}.  There are three temperatures defined
in the disc, the midplane temperature, $T_{\rm c}$, the surface layer
temperature, $T_{\rm m}$, and the surface temperature, $T_{\rm
  e}$. These are related to each other by considering the energy
balance in a layered disc model in thermal equilibrium to find
\begin{equation}
T_{\rm m}^4 = \tau_{\rm m} T_{\rm e}^4
\label{t1}
\end{equation}
and 
\begin{equation}
\sigma T_{\rm c}^4= \frac{9}{8}\Omega^2 \left( \nu_{\rm m} \Sigma_{\rm
  m}\tau_{\rm m}+\nu_{\rm d}\Sigma_{\rm d}\tau \right)
\label{t2}
\end{equation}
\citep{martin11a}, where $\tau_{\rm m}$ and $\tau_{\rm d}$ are the
optical depths of the magnetic surface layers and dead zone
respectively and $\tau=\tau_{\rm m}+\tau_{\rm d}$.  We have found that
the effects of irradiation from the central star are negligible for
the outburst model presented here. The characteristic temperature for
outbursts is of order $T_{\rm crit} = 800\,\rm K$. Stellar irradiation
cannot produce such high mid-plane disc temperatures on the AU scale
where GM outbursts occur.  By solving equations~(\ref{mdot})
and~(\ref{energy}) with~(\ref{t1}) and~(\ref{t2}) we find the steady
state disc structure. In the following two sections we analyse such
solutions.

\section{Fixed $\alpha_{\rm \lowercase{d}}$ in the Dead Zone}
\label{sec:ad}

Here we consider steady state disc solutions with a constant residual
viscosity in the dead zone, $\alpha_{\rm d}=\,\rm constant$. This may
be appropriate for modelling discs with a dead zone that is unstable
to hydrodynamic instabilities, such as the baroclinic instability, for
example. 

A disc with a dead zone with zero residual viscosity can only be in
steady state if the disc is sufficiently massive to be self
gravitating.  However, with an additional small viscosity in the dead
zone, further steady state solutions are possible if the dead zone and
active layer together can transport material at a rate equal to the
infall accretion rate.  Consider the case of a disc that has a very
small level of residual turbulence (very small $ \alpha_{\rm d}$). In
order to carry the required mass flux, the dead zone surface density
must be very high. As a result, the disc optical depth and therefore
midplane temperature must be high. However, the midplane disc
temperature of such a steady state must be less than $T_{\rm crit}$,
the critical temperature for the onset of MRI. Otherwise, the MRI disc
turbulence would set in. The resulting mass flux would then be too
high for a steady state and the disc would undergo outbursts by
cycling between these levels of turbulence. Hence, a steady state disc
does not exist for all dead zone turbulence levels $\alpha_d$. Hence,
the steady solution does not exist for all disc parameters in the GM
unstable region of the disc.

\subsection{Locally Steady Models}

We first consider a disc model with $\Sigma_{\rm crit}=200\,\rm
g\,cm^{-2}$.  There are four steady state disc solutions; the
thermally ionised MRI, the externally ionised MRI, the GM and the
viscous dead zone solution.  The first three of these solutions are
independent of $\alpha_{\rm d}$, as discussed in the previous
section. The top left plot in Fig.~\ref{data} shows where each of
these three steady solutions exist for varying infall accretion rate
and radius. The unshaded region shows the parameter space for which
the disc is unstable to the GM instability with $\alpha_{\rm
  d}=0$. Note that we do not include the thermal-viscous instability
in this work.  \cite{zhu07} found that in order to fit the Spitzer
Space Telescope IRS spectrum of the eponymous outbursting system FU
Ori, the rapidly accreting, hot inner disk must extend out to about $1
\,\rm AU$, inconsistent with a pure thermal instability model.  We
therefore concentrate on the GM instabilities that occur on the AU
scale.

In the remaining three plots we show at radii $R=1$, $3$ and $5\,\rm
AU$ where the four steady solutions exist for varying infall accretion
rate and dead zone viscosity. For the viscous dead zone solution, we
find the critical $\alpha_{\rm d}$ for which a steady solution exists
by setting $T_{\rm c}=T_{\rm crit}$. This is shown in the thick
lines. In the dark shaded regions we show the range of $\alpha_{\rm
  d}$ for which a steady viscous dead zone solution exists (with
$T_{\rm c}<T_{\rm crit}$ and $Q>Q_{\rm crit}$).  In the dark shaded hatched
region (shown at $R=1$ and $R=3\,\rm AU$) the accretion rate through
the dead zone is smaller than that through the active layer. In plots
that show no hatched region there is no steady viscous dead zone
solution that has an accretion rate through the dead zone less than
that through the active layer.  Especially for large radii, and high
accretion rates, the viscous dead zone solution requires a large
viscosity in the dead zone for a steady state to exist.

In Fig.~\ref{data2} we show the same plots for a disc with a smaller
active layer of $\Sigma_{\rm crit}=20\,\rm g\,cm^{-2}$. With smaller
active layer surface density there is a larger range of GM unstable
accretion rates and similarly a slightly larger range of $\alpha_{\rm
  d}$ for which a steady solution exists. However, the region in which
a steady viscous dead zone solution exists, with $\dot M_{\rm d}<\dot
M_{\rm m}$, is significantly smaller. 

We have also investigated the effects of changing various parameters.
The functional form for $\alpha_{\rm g}$ has little effect on the disc
stability \citep[see][for a discussion]{martin13}. The value of the
critical Toomre parameter below which graviataional instabilities
operate, $Q_{\rm crit}$, may be as small as 1.7 \citep[see
  e.g.][]{boley06,durisen07}. However, we find the effects of such a
change to be negligible to the work presented here. The critical
temperature above which the MRI operates, $T_{\rm crit}$, on the other
hand does change the results quantitatively. For example, the
instability region described in the top left plot of Fig.~1 is shifted
a small amount when we double $T_{\rm crit}$. However, qualitatively,
this does not affect our conclusions on the disc stability.

\subsection{Local Limit Cycle}

We consider the limit cycle in the $\Sigma-\dot M$ plane described in
\cite{martin11a} . This limit cycle explains the GM disc instability
as transitions between steady state solutions when the infall
accretion rates lies in a region where there is no steady state
solution. Here, we include the additional steady state viscous dead
zone solution.

The surface temperature is related to the accretion rate through
equations~(\ref{mdot}) and~(\ref{energy}) with
\begin{equation}
T_{\rm e}=\left(\frac{3\dot M\Omega^2}{8 \pi \sigma}\right)^\frac{1}{4}.
\end{equation}
The disc has $\Sigma_{\rm crit}=200\,\rm g\,cm^{-2}$ and we show a
radius $R=3\,\rm AU$. The steady state solutions with $\alpha_{\rm
  d}=0$ \citep[from][]{martin11a} are shown in the thick black
lines. Steady state solutions with $\alpha_{\rm d}>0$ are shown in the
grey/white lines for three different values of $\alpha_{\rm d}$. The
white part shows where the flow through the dead zone is larger than
that through the active layer and the grey part shows the solutions
where the flow through the dead zone is smaller than that through the
active layer. The upper ends the of the white lines are where $T_{\rm
  c}=T_{\rm crit}$. In the case of $\alpha_{\rm d}=10^{-5}$, the grey
line ends where the solution becomes self gravitating and there is no
white part.

If there remains a range of accretion rates for which no steady
solution exists, the disc will be unstable. As shown in
\cite{martin11a}, if the accretion rate on to the disc lies in the
unstable range, the disc will exhibit a limit cycle as it transitions
between the two steady states. For the gravo-magneto limit cycle (with
zero viscosity in the dead zone), the two steady states are the
thermal MRI and the self-gravitating solution. However, with a non
zero $\alpha_{\rm d} > 10^{-5}$ in the dead zone, a similar limit
cycle exists where the disc transitions between the thermal MRI and
the viscous dead zone steady state. The viscosity in the dead zone
drives extra turbulence and heating within the dead zone that
eventually triggers the thermal MRI and an outburst ensues.

In order to stop the outburst cycle, the steady state viscous dead
zone solutions would need to cover the whole of the unstable
region. Thus, to stabilise the disc to the GM disc instability
requires a rather high dead zone viscosity, $\alpha_{\rm d}>
10^{-3}$. In this case, most of the accretion flow is through the dead
zone, rather than the active layers.

\begin{figure}
\includegraphics[width=8.4cm]{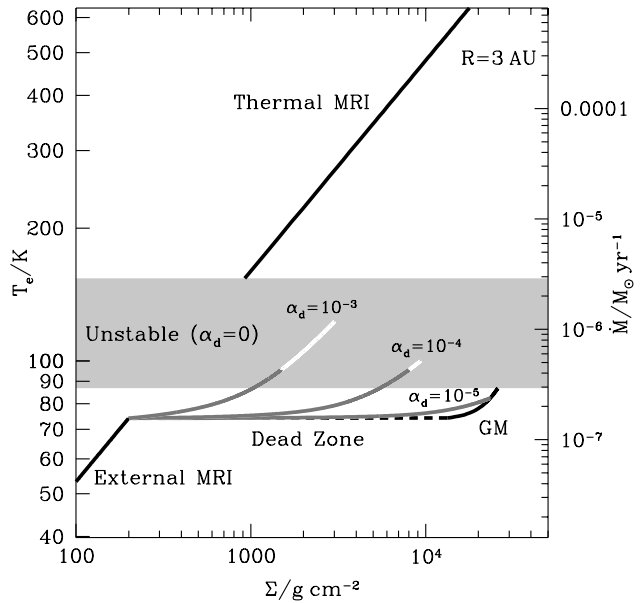}
\caption{$\Sigma-\dot M$ or $T_{\rm e}$ plane at $R=3\,\rm AU$ for a
  disc with $\Sigma_{\rm crit}=200\,\rm g\,cm^{-2}$. The thick black
  lines show the steady state solutions with no viscosity in the dead
  zone. These are the thermal MRI, external MRI and GM solutions. The
  dashed line shows where there is a dead zone solution with
  $\alpha_{\rm d}=0$, that is not steady. The shaded region shows the
  GM unstable region with $\alpha_{\rm d}=0$.  The grey/white lines
  show steady state viscous dead zone solutions for $\alpha_{\rm
    d}=10^{-3}$, $10^{-4}$ and $10^{-5}$. The grey part denotes where
  $\dot M_{\rm d}<\dot M_{\rm m}$ and the white part where $\dot
  M_{\rm d}>\dot M_{\rm m}$. }
\label{scurve}
\end{figure}

\subsection{Globally Steady Models}

Here we consider the range of accretion rates for which a steady
solution exists at all radii in the disc.  In \cite{martin13} we found
the radius of the inner edge of the GM unstable region to be
\begin{equation}
R_{\rm crit}= 1.87 \, \frac{ \dot M'^{4/9} M'^{1/3}}{\alpha'^{2/9} T_{\rm
   crit}'^{14/15}} \,\rm AU
\end{equation}
where $\dot M'=\dot M/10^{-6}\,\rm M_\odot\,yr^{-1}$,
$\alpha'=\alpha/0.01$, $M'=M/\rm M_\odot$ and $T_{\rm crit}'=T_{\rm
  crit}/{800\,\rm K}$. This is the radius where $T_{\rm c}=T_{\rm
  crit}$ for the fully MRI active solution and is shown as the line
under which a steady thermally ionised MRI solution exists in the top
left plot in Fig.~\ref{data}. Outside of this radius, there is not a
thermally ionised MRI steady state solution.

If there was no limit on the temperature of a steady viscous dead zone
solution (it must satisfy $T_{\rm c}<T_{\rm crit}$, otherwise the MRI is
triggered), we could find a steady viscous dead zone solution for all
disc parameters. The solutions would fill the unstable region in the
top left plot of Fig.~1. Because the temperature of the disc decreases
with radius, if there is a viscous dead zone solution at $R=R_{\rm
  crit}$ with $T_{\rm c}<T_{\rm crit}$, then there is a solution throughout
the unstable region. In this case, there is no unstable region and
there is a steady solution throughout the disc. Similarly, if there is
another type of steady solution at $R=R_{\rm crit}$ (an externally
ionised MRI or GM steady solution), then there is no GM unstable
region and there is a steady solution throughout the disc.  This can
be seen in Fig.~1.  At $R=R_{\rm crit}(\dot M)$, if
$\Sigma<\Sigma_{\rm crit}$ then the disc has an externally ionised
fully MRI active steady solution. This occurs for $\dot M<3.01\times
10^{-8}\,\rm M_\odot\,yr^{-1}$.  If the accretion rate is sufficiently
high ($\dot M>1.10\times 10^{-4}\,\rm M_\odot\,yr^{-1}$) there is a GM
steady state solution at $R=R_{\rm crit}(\dot M)$. For accretion rates
between these two values, we determine possible GM stable discs that
result from having a nonzero $\alpha_d$.

\begin{figure*}
\includegraphics[width=8.4cm]{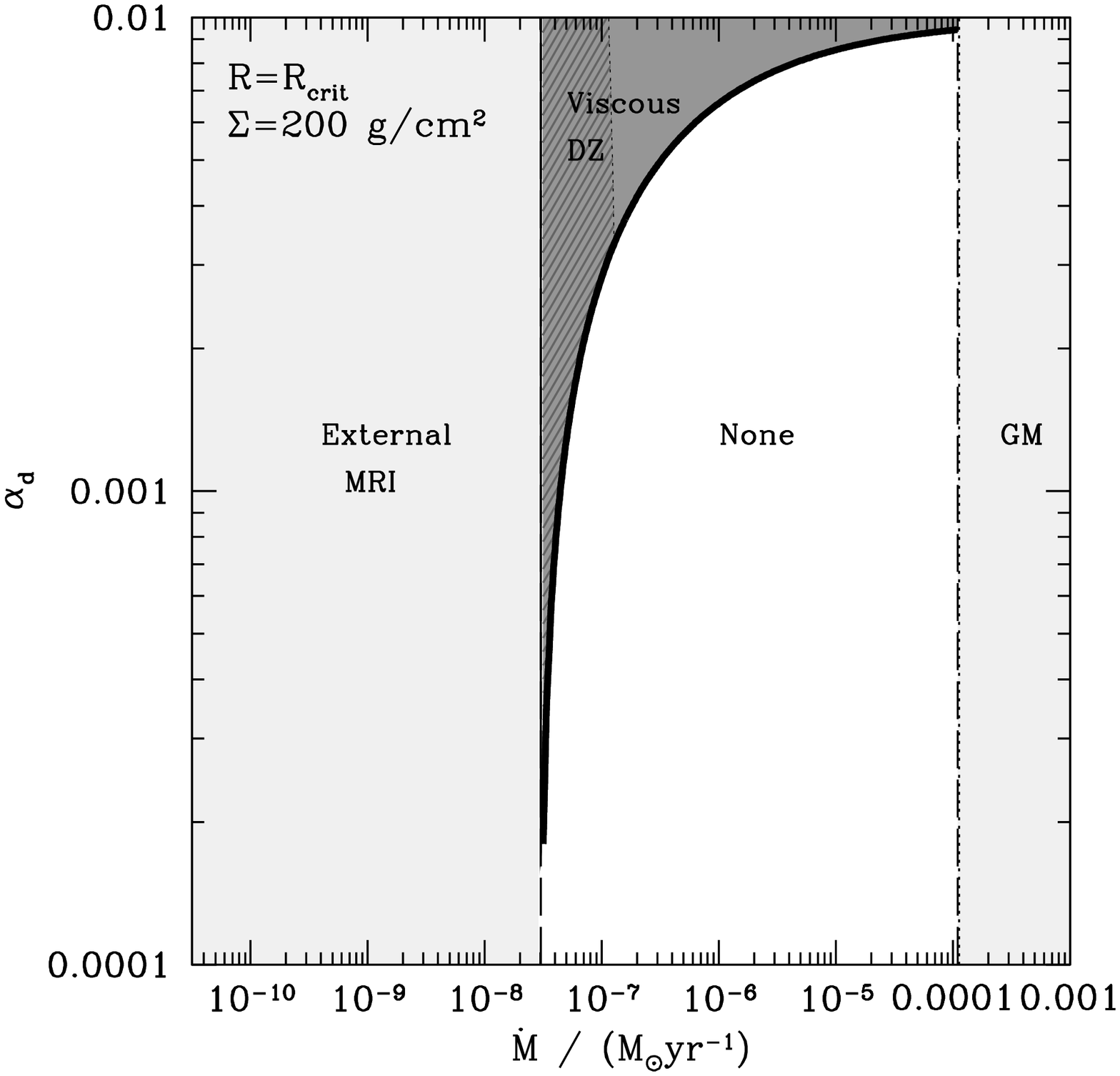}
\includegraphics[width=8.4cm]{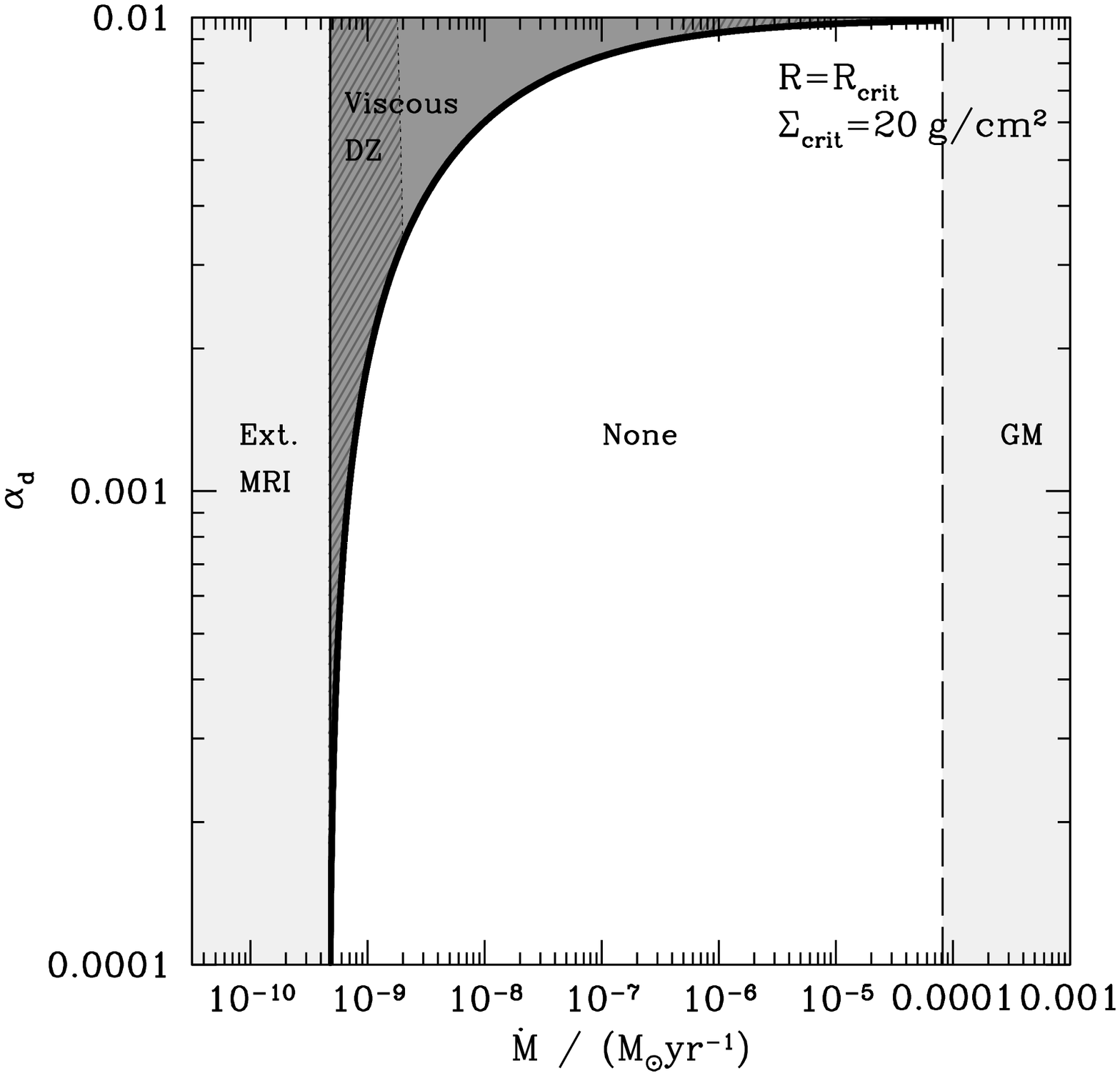}
\caption{The globally steady disc solutions for $\Sigma_{\rm
    crit}=200\,\rm g\,cm^{-2}$ (left) and $\Sigma_{\rm crit}=20\,\rm
  g\,cm^{-2}$ (right). (Same as the top right panels in
  Fig.~\ref{data} and Fig.~\ref{data2}, for example, but with $R=R_{\rm
    crit}(\dot M)$.) }
\label{changer}
\end{figure*}

The critical $\alpha_{\rm d}$ above which a steady viscous dead zone
solution occurs where $T_{\rm c}=T_{\rm crit}$.  We show where the
steady state solutions exist in Fig.~\ref{changer}. This is similar to
the plots in Figs.~\ref{data} and~\ref{data2} but the radius is not
constant with accretion rate, we take $R=R_{\rm crit}(\dot M)$.  This
represents the globally steady disc solutions. A globally steady disc
solution can be found for all accretion rates if $\alpha_{\rm d}$ is
very close to $\alpha_{\rm m}$. For most accretion rates, the dead
zone is transporting the majority of the flow, even for the high
active layer surface density. For a very small range of accretion
rates, a globally steady disc is possible with $\alpha_{\rm d}\approx
10^{-3}$.

\subsection{Summary}

If the viscosity in the dead zone is driven by hydrodynamic
instabilities such as the baroclinic instability, there is no
requirement on the accretion flow through the dead zone being less
than that through the active layer. In this section we showed that a
high value of $\alpha_{\rm d}\gtrsim 10^{-3}$ is needed to stabilise
the disc to outbursts. The amount of turbulence that can be generated
by hydrodynamic instabilities is unlikely to be this high. If the
baroclinic instability can drive turbulence, it will be significantly
smaller than turbulence generated by the MRI
\citep[e.g.][]{armitage11}. Thus we find it is unlikely that
hydrodynamic instabilities in the dead zone can stabilise the disc to
the outburst behaviour. However, outbursts may be triggered by heating
from the viscosity in the dead zone, rather than the heating from
self-gravity.

\section{Fixed fractional accretion rate in the dead zone}
\label{sec:md}

In this section we consider steady state solutions where $\alpha_{\rm
  d}$ is not a fixed number but the accretion rate through the dead
zone is a constant fraction, $f$, of that through the active layer,
\begin{equation}
\dot M_{\rm d}=f\dot M_{\rm m}. 
\label{mdot2}
\end{equation}
This is equivalent to
\begin{equation}
\alpha_{\rm d}=\alpha_{\rm m} \min \left(1,f \frac{T_{\rm m}}{T_{\rm
    c}}\frac{\Sigma_{\rm m}}{\Sigma_{\rm d}}\right).
\end{equation}
The viscosity is limited so that a larger $\alpha$ is not driven in
the dead zone compared to that in the active layer even when the dead
zone layer has a very small surface density.

This viscosity prescription is applicable to dead zone viscosities
driven by hydrodynamic waves that are excited in the dead zone by the
turbulence from the active layer. Thus, its magnitude depends upon the
depth of the active layer compared to the dead zone layer
\citep{fleming03}. The MHD calculations in \cite{turner08} find that
the dead zone transports material at a rate with fraction, $f$, in the
range $0.04$ to $0.61$ depending on disc parameters such as the
presence of dust grains and the ionising sources.  \cite{simon11} find
a more a more constant value of $\alpha_{\rm d}=10^{-4}$ in the dead
zone but consider only radii of $4\,\rm AU$ and $10\,\rm AU$ in a
minimum mass solar nebula (MMSN). Similarly, all of these calculations
assume a MMSN surface density that may be much smaller than that in a
real time-dependent disc. Thus, in this work we consider a range of
values for $f$.

\begin{figure}
\includegraphics[width=8.4cm]{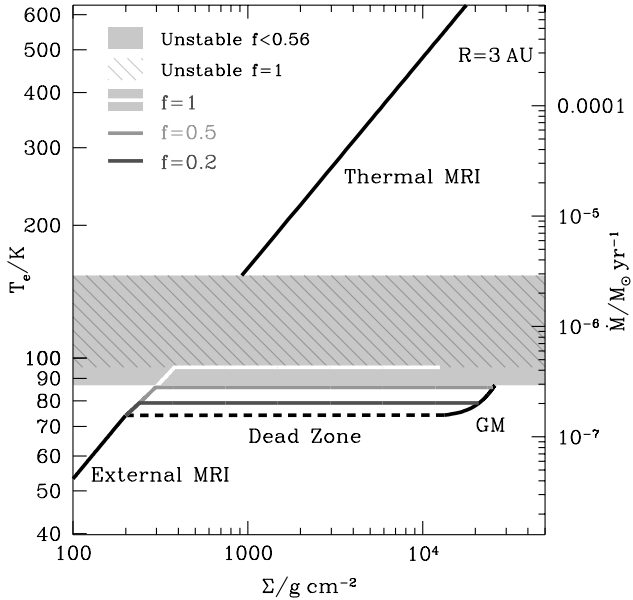}
\caption{$\Sigma-\dot M$ or $T_{\rm e}$ plane at $R=3\,\rm AU$ for a
  disc with $\Sigma_{\rm crit}=200\,\rm g\,cm^{-2}$. The thick black
  lines are the same as in Fig.~\ref{scurve}, the steady state
  solutions with $\alpha_{\rm d}=0$. The remaining thick lines show
  steady state solutions with $\alpha_{\rm d}$ defined such that $\dot
  M_{\rm d}=f\dot M_{\rm m}$ with $f=0.2$ (dark grey), $f=0.5$ (light
  grey) and $f=1$ (white). The grey shaded area shows the unstable
  region for $f<0.56$. The hatched area shows the unstable region for
  $f=1$. }
\label{scurve2}
\end{figure}

In Fig.~\ref{scurve2} we show the local limit cycle for such solutions
with different $f$ at a radius $R=3 \,\rm AU$.  As the surface density
of a steady solution increases, the viscosity in the dead zone
decreases. Thus, on the far right of the diagram, the GM branch is not
noticeably affected by the addition of a dead zone viscosity. Because
a larger amount of material flows through the disc compared to the
zero viscosity dead zone solution, the solutions form a branch above
the dead zone branch.  For $f<0.56$, the new viscous dead zone branch
connects to the GM branch on the right hand side and the range of GM
unstable accretion rates is unaffected. The outburst will be
  triggered by the heating from self gravity.  However, for $f>0.56$,
the branch is above the top of the GM branch and ends where $T_{\rm
  c}=T_{\rm crit}$. The range of unstable accretion rates is slightly
decreased, as shown in the hatched region for $f=1$ compared to the
shaded region for $f=0$. In this case, the outburst will be
  triggered by the heating from the viscosity in the dead zone rather
  than the heating from self-gravity.

With the flow through the dead zone restricted to be smaller than that
through the active layer ($f<1$), the additional steady states do not
decrease the range of unstable accretion rates significantly.  The
dead zone viscosity can only produce an increase in the lowest
unstable accretion rate by a fraction of the accretion rate through
the active layer.  To prevent the outburst cycle, the steady state
viscous dead zone solutions would need to cover the whole of the
unstable region. Thus we find that even with these additional steady
state solutions, closure of the gap is not likely unless the gap is
small already with $\alpha_{\rm d}=0$. Thus, hydrodynamic turbulence
in the dead zone, generated from the turbulence in the active layers
is unlikely to stabilise a disc to the GM disc instability.

\section{Conclusions}

With zero viscosity in the dead zone, for a range of accretion rates,
a disc is unstable to the GM disc instability (where the turbulence
cycles from gravitationally produced to magnetically produced). We
have examined a further class of disc solutions where the dead zone
has a small residual viscosity that allows the possibility of a steady
state.  We have considered two types of turbulence in the dead
zone. First, we took a constant $\alpha$-viscosity. This may be
applicable to hydrodynamic instabilities such as the baroclinic
instability. We found that a steady solution may only be found if the
turbulence in the dead zone is very high, comparable to that in the
MRI active layers, but this is unlikely. The second type of turbulence
we considered generates an accretion flow through the dead zone that
is a fixed fraction of that through the MRI active layers. This may be
applicable to hydrodynamic turbulence generated from the MRI active
layers.  The range of unstable accretion rates is hardly affected
by the additional dead zone viscosity. However, the triggering
mechanism for the outbursts may be the heating from the viscosity in
the dead zone, rather than the self-gravity. Thus, we find it is
unlikely that either type of turbulence within the dead zone can
stabilise a disc and prevent the outbursting behaviour.

\section*{Acknowledgements}

We thank an anonymous referee and Jacob Simon for useful
comments. RGM's support was provided under contract with the
California Institute of Technology (Caltech) funded by NASA through
the Sagan Fellowship Program. SHL acknowledges support from NASA grant
NNX11AK61G.


\label{lastpage}
\end{document}